\documentclass[journal]{IEEEtran}

\usepackage{graphicx}
\usepackage{makecell}
\usepackage{algorithm} %format of the algorithm 
\usepackage{algorithmic} %format of the algorithm
\usepackage{color} 
\usepackage{multirow} %multirow for format of table 

\usepackage{threeparttable}

\ifCLASSINFOpdf
\else
\fi
\begin{document}
\title{ Large-scale and High-speed Privacy Amplification for FPGA-based Quantum Key Distribution}

\author{Yan Bingze, Li Qiong, Mao Haokun% <-this % stops a space
\thanks{M. Shell was with the Department
of Electrical and Computer Engineering, Georgia Institute of Technology, Atlanta,
GA, 30332 USA e-mail: (see http://www.michaelshell.org/contact.html).}% <-this % stops a space
\thanks{J. Doe and J. Doe are with Anonymous University.}% <-this % stops a space
\thanks{Manuscript received April 19, 2005; revised August 26, 2015.}}


\markboth{Journal of \LaTeX\ Class Files,~Vol.~14, No.~8, August~2015}%
{Shell \MakeLowercase{\textit{et al.}}: Bare Demo of IEEEtran.cls for IEEE Journals}



\maketitle

\begin{abstract}
The FPGA-based Quantum key distribution (QKD) system is an important trend of QKD systems. It has several advantages, real time, low power consumption and high integration density. Privacy amplification is an essential part in a QKD system to ensure the security of QKD. Existing FPGA-based privacy amplification schemes have an disadvantage, that the throughput and the input size of these schemes (the best scheme $116Mbps$@$10^6$) are much lower than these on other platforms (the best scheme $1Gbps$@$10^8$). This paper designs a new PA scheme for FPGA-based QKD with multilinear modular hash-modular arithmetic hash (MMH-MH) PA and number theoretical transform (NTT) algorithm. The new PA scheme, named large-scale and high-speed (LSHS) PA scheme, designs a multiplication-reusable architecture and three key units to improve the performance. This scheme improves the input size and throughput of PA by above an order of magnitude. The throughput and input size of this scheme ($1Gbps$@$10^8$) is at a comparable level with these on other platforms. 
\end{abstract}


% \fi
\IEEEpeerreviewmaketitle

\section{Introduction}
% PA简介

% QKD背景介绍
\IEEEPARstart{Q}{uantum} key distribution (QKD) is a notable technique which exploits the principle of quantum mechanics to perform the information theoretical security key distribution between two remote parties, named Alice and Bob \cite{BennettCharlesandBrassard1984}. A QKD system can be divided into two parts, the quantum optical subsystem and the postprocessing subsystem. The quantum optical subsystem is for the preparation, transmission and measurement of quantum states. The postprocessing subsystem is to complete the correctness and security of the final secure key \cite{Mao2019}. A Field-Programmable-Gate-Array (FPGA) based QKD system means that the control part of its quantum optical subsystem and its postprocessing subsystem is implemented by a FPGA~\cite{Zhang2012a,Constantin2017c}. The advantages of a FPGA-based QKD system are real time, low power consumption, high integration density. A FPGA-based QKD system can be combined with the integrated optical circuit to implement the QKD system on chip, which will provide improved performance, miniaturization and enhanced functionality of the QKD system\cite{Sibson2017a}. 

Privacy amplification is a necessary part in quantum key distribution~\cite{Bennett1995}. It is the art of distilling a highly secure key from a partially secure string by public discussion between two parties. It is one of the main bottlenecks of the FPGA-based QKD system. 

%% PA目前最关键的问题是？

The lacking input block size is the most critical problem of PA for a FPGA-based QKD system. The input block size of PA has significant impact on the final key rate of QKD system~\cite{Furrer2012,Tomamichel2012}. The largest input block size of existing FPGA-based PA schemes is $10^6$~\cite{Li2019a}, while the common input block size of PA schemes on other platforms is more than $10^8$~\cite{Tang2019,XiangyuWangYichenZhangSongYu2016a}. 

%% 主线任务就是要提高码长
%先分析原因

The constricted computing resource of FPGA is the main reason of lacking the input block size of FPGA based PA. 

%再介绍已有的提升码长的方法

To realize large input block size with the constricted computing resource, a few schemes have been proposed on other platforms such as the length-compatible PA on the GPU~\cite{XiangyuWangYichenZhangSongYu2016a}, the HiLS PA on the CPU~\cite{Tang2019} and the MMH-MH PA on the CPU. The length-compatible PA and the HiLS PA are both based on Toeplitz-hash PA algorithm. They take the advantage of Toeplitz-hash to improve input block size by dividing the long input sequence into short block. It is convenient to design a similar scheme on FPGA according to these schemes, because there are already two methods to implement Toeplitz-hash PA on FPGA, which are block parallel method and FFT-based method. However, we do not regard these Toeplitz-based methods as the most suitable method to design a large input block size PA scheme on FPGA.The unsuitability of the block parallel method is that it is hard to overcome the lack of real-time, because its computation complexity is as high as $\mathop{O}(n^2)$. The unsuitability of the FFT-based method is that it relies on the floating-point arithmetic, which may bring the calculation error and impact on security of key. Floating-point arithmetic also increases the memory consumption and requires external storage, which affects the integration level of the system.

The Multilinear Modular Hash - Modular Arithmetic Hash (MMH-MH) PA is a new PA algorithm that can realize large input block size PA with the constricted computing resource~\cite{Bingze2021}. It can be implemented by number theory transform (NTT) and provides strong real-time with the $\mathop{O}(n\log n)$ computation complexity. NTT uses integer arithmetic instead of floating-point arithmetic, which avoids the calculation error and external storage. Therefore, a large scale PA scheme based on the MMH-MH PA algorithm is designed in this paper to improve the performance of the FPGA-based QKD system.

% 在本文中，首先介绍了MMH-MH PA算法的具体过程及其安全性分析（第二章）。可以看到MMH-MH PA算法的两大部分 MMH和MH的核心操作都是大数乘法，大数乘法又是一种资源消耗很大的运算，因此本文首先为方案设计了一种大数乘法可复用的总体结构，并设计了相应的控制单元，随后介绍了方案中一些重要参数的确定方法，包括如何根据压缩比来获得最大的输入码长以及如何根据输入密钥的速率来优化计算资源；随后本文设计了方案中的各个关键模块：1.一个基于NTT的大数乘法运算核，其能够完成768Kb数的大数乘法运算；2. 一个节省资源且高效的模累加运算模块；3.一个流水线式的二进制取模模块（第三章）。最后，我们对该方案的性能进行了测试，我们首先计算了不同压缩比下，该方案所能达到的输入码长，该方案在0.3压缩比下，码长可以达到2Mb；0.1压缩比下码长可达到7Mb；0.01压缩比下码长可达到70Mb。随后我们分别以典型DV-QKD系统和典型CV-QKD系统为例，实验了该方案在随通信距离的变化对码长的影响，可以发现该方案相较于已有FPGA方案能够提高最终成码率，并且在通信距离较长的情况下，提升效果更为明显。再然后我们对吞吐率进行了测试，测试结果显示，该方案具有极好的实时性，吞吐率远高于已有方案甚至可以超过GPU实现方案，非常适合高速的QKD系统。最后我们对比了我们方案和已有方案的资源消耗，可以看到我们的方案避免了对于外存的依赖，并且在计算资源上可以通过参数调整降得很低，但是该方案的片内存储资源消耗是相对较高的，因此更适合片内存储资源丰富的芯片使用。

The principle and security analysis of MMH-MH PA Algorithm is introduced as a basis for this work in section 2. The multiplication of large numbers is the major part in both multilinear modular hash and modular arithmetic hash. Therefore, a multiplication-reusable structure and its control unit for the MMH-MH PA is designed. Subsequently, the optimization method of the PA input size according to the compression ratio is introduced. The design of three key units in this scheme is introduced: 1. the NTT-based multiplication unit, which can accomplish 768Kb sized multiplication operation; 2. the memory-saving and efficient modular accumulation unit; 3. the pipelined binary modulo unit. The design of scheme is all introduced in section 3. The performance of our scheme is evaluated in section 4. The input block size at different compression ratio $R_{PA}$ is calculated. The input block size of our scheme can be $2\time 10^6$ at $R_{PA}=0.3$, $7\times 10^6$ at $R_{PA}=0.1$ and $7 \times 10^7$ at $R_{PA}=0.01$. The influence of our PA scheme on the system final key rate as the transmission distance changes is simulated according to the key parameters of a typical DV-QKD system and a typical CV-QKD system. The results indicate that our schemes can improve the final key rate compared with existing FPGA-based PA scheme, and the improvement is more obvious in the CV-QKD system and long transmission distance situation. The throughput of our scheme is evaluated at different input block size. The results reveal that The throughput of our scheme improves an order of magnitude compared with existing FPGA-based schemes and our scheme is right for system with high demand of real-time. The resource consumption of our scheme is assessed and compared with existing schemes. The computation resource of our scheme can be optimized based on the demand of throughput, and it can be pretty low with low demand of real-time. Our scheme is freed from the dependence on external storage, while still costs relatively more internal storage, about half of available storage on our chip. 
%%这句作结论有点奇怪。。
%%可能后面需要补充一些结论

\section{MMH-MH PA Algorithm}

The multilinear modular hashing-modular arithmetic hashing (MMH-MH) PA algorithm is introduced in this section. It is the fundamental of this scheme.  

%\subsection{Privacy Amplification}

%\subsection{MMH-MH PA Algorithm}
%经典PA算法的主要过程是使用通用哈希族中的随机函数对输入序列进行压缩。MMH-MH PA 则进行两次压缩。算法的主要流程按图1所示。

% Fig 1
\begin{figure}
	\includegraphics[height=7.2cm,width=9cm]{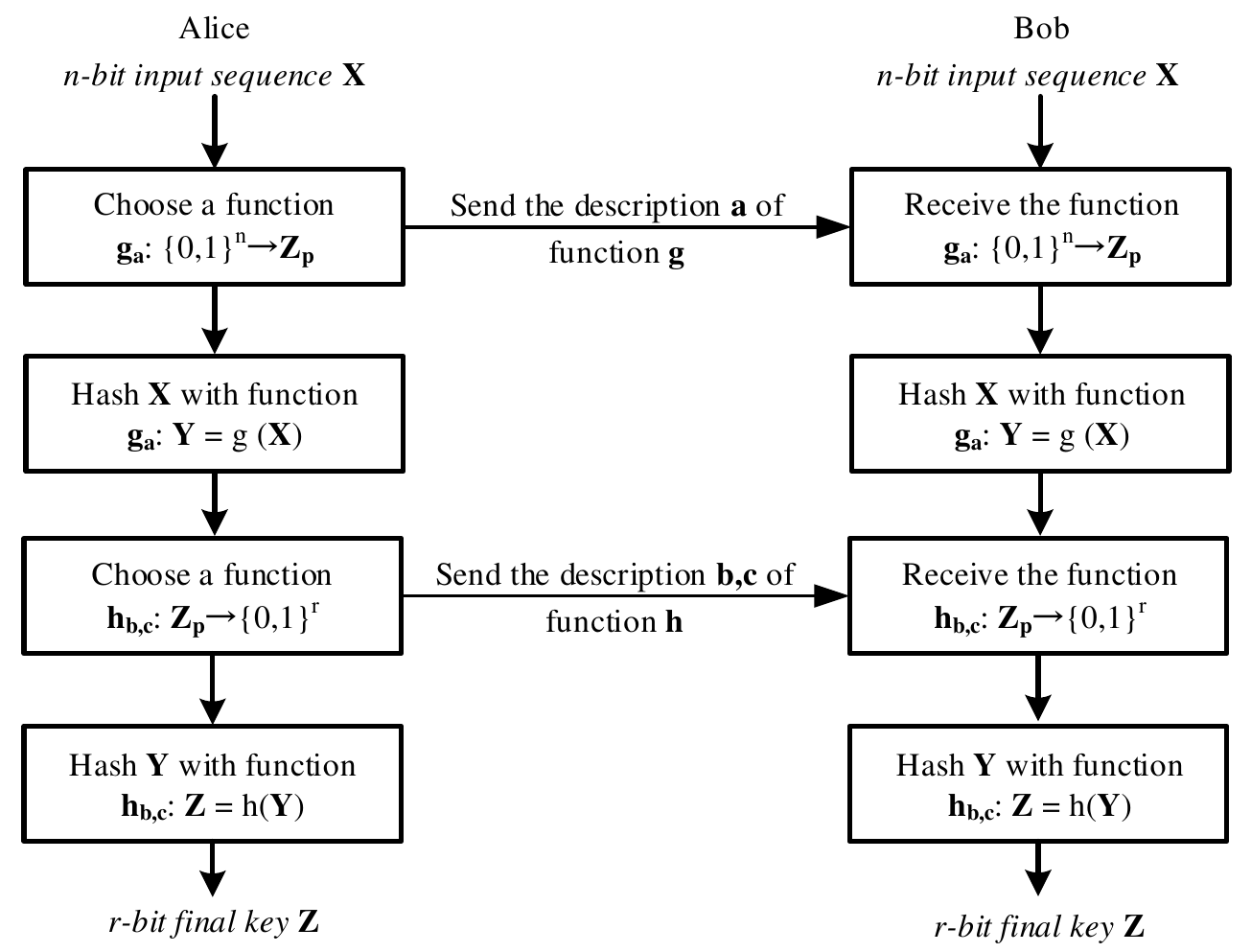}
	\caption{the main steps of MMH-MH PA algorithm}
	\label{fig:1}       % Give a unique label
\end{figure} 

The main process of a typical PA algorithm is compressing the input sequence with a hash function randomly chosen from the universal hash family. The MMH-MH PA algorithm performs the compression twice with different universal hash families instead of once. The main steps of MMH-MH PA algorithm are indicated as Fig. \ref{fig:1}, where $n$ is the length of input sequence. $r$ is the length of final key. The definition of multilinear modular hash and modular arithmetic hash are given as follow.

\paragraph{Definition of Multi-linear Modular Hashing}
Let $p$ be a primer and let $k$ be an integer $k > 0$. Define a family multi-linear modular hashing of functions from $Z^k_p$ to $Z_p$ as follows:
\begin{equation}{\rm{MMH}}: = \left\{ {{{\mathop{\rm g}\nolimits} _a}:Z_p^k \to {Z_p}\left| {a \in Z_p^k} \right.} \right\}\end{equation}
where the function ${{{\mathop{\rm g}\nolimits} _a}}$ is defined for any $a = \left\langle {{a_1}, \cdots ,{a_k}} \right\rangle $, $x = \left\langle {{x_1}, \cdots ,{x_k}} \right\rangle$, ${a_i},{x_i} \in {Z_p}$,
\begin{equation}{{\mathop{\rm g}\nolimits} _a}\left( x \right): = a \cdot x\bmod p = \sum\limits_{i = 1}^k {{a_i}{x_i}\bmod p} \end{equation}

MMH family is an universal hashing family~\cite{Carter1979a}, its collision probability $\delta$ is $1/|Z_p|$, and the proof can be found in~\cite{Halevi1997}.

\paragraph{Definition of Modular Arithmetic Hashing}
Let $\alpha$ and $\beta$ be two strictly positive integers, $\alpha > \beta$. Define a family modular arithmetic hashing of functions from $2^\alpha$ to $2^\beta$ as follows:
\begin{equation}{\rm{MH}}: = \left\{ {{h_{b,c}}:{Z_{{2^\alpha }}} \to {Z_{{2^\beta }}}\left| {b,c \in {Z_{{2^\alpha }}}} \right.,\gcd (b,2) = 1} \right\}\end{equation}
where the function $h_{b,c}$ is defined as follows:
\begin{equation}{h_{b,c}}(x): = {{\left( {b \cdot x + c\bmod {2^\alpha }} \right)} \mathord{\left/
		{\vphantom {{\left( {b \cdot x + c\bmod {2^\alpha }} \right)} {{2^{\alpha  - \beta }}}}} \right.
		\kern-\nulldelimiterspace} {{2^{\alpha  - \beta }}}}\end{equation}

Modular Arithmetic Hashing can be designed for PA algorithm itself, while it can not split the input and handle it separately. the output set of modular arithmetic hashing is variable length bit sequence. So it can be combined with MMH to design a new PA algorithm. 

The specific process of the MMH-MH PA algorithm is given as Algorithm 1. In details, the prime number $p$ is suggested to be a Mersenne prime. The form of a Mersenne prime is $M_\gamma = 2^\gamma - 1$. The length of input sequence is $n = \gamma \times k$. $x_i = 2^{\gamma}-1$ is a special case, the data $x_i = 2^\gamma - 1$ should be cast away and reload. 

%算法的规范化表示

\begin{algorithm} %算法开始 
	\caption{MMH-MH PA algorithm} %算法的题目 
	\label{alg1} %算法的标签 
	\begin{algorithmic}[1] %此处的[1]控制一下算法中的每句前面都有标号 
		\REQUIRE Input Data: $x \in Z_{2^{k\times\gamma}}$. \\\quad\quad Random numbers:$a\in Z^k_p$, $b,c \in Z_{2^\gamma}$, $\gcd (b,2) = 1$.  \\\quad\quad//$p=M_\gamma=2^\gamma - 1$ 
		\ENSURE $z \in Z_{2^\beta}$ //$\gamma > \beta$%输出结果(此处的ENSURE默认关键字为Ensure在上面已自定义为Output) 
		
		\STATE $x = \left\langle {{x_1}, \cdots ,{x_k}} \right\rangle$ //split data $x$
		\STATE $a = \left\langle {{a_1}, \cdots ,{a_k}} \right\rangle$ //split data $a$
		\IF{$x_i = 2^\gamma - 1$($i=1,...,k$)} 
		\STATE break;   //Reload data $x_i$
		\ELSE
		\FOR{$i=0$ to $k$}
		\STATE $y_i=a_i\times x_i$  
		\ENDFOR 
		\STATE $y = \sum\limits_{i = 1}^k {{y_i}\bmod p} $  \quad/*MMH function: $y = {{\mathop{\rm g}\nolimits} _a}(x)$*/
		\STATE $z = {{\left( {b \cdot y + c\bmod {2^\alpha }} \right)} \mathord{\left/
				{\vphantom {{\left( {b \cdot x + c\bmod {2^\alpha }} \right)} {{2^{\alpha  - \beta }}}}} \right.
				\kern-\nulldelimiterspace} {{2^{\alpha  - \beta }}}}$ \\\quad\quad/*MH function: $z = {{\mathop{\rm h}\nolimits} _{b,c}}(y)$*/
		\ENDIF
	\end{algorithmic} 
\end{algorithm}

 Because the process of MMH-MH PA is different from that of traditional PA algorithms, we have proven that the security of MMH-MH PA is similar with other PA algorithms in \cite{Bingze2021}. MMH-MH PA algorithm requires an additional condition to guarantee the security, that is the length of final key $r$ should be much less than $\gamma$, specifically $r<\gamma-s$ ($s$ is the information theory security parameter of QKD).

%\subsection{Security of MMH-MH PA}

\section{Large-Scale and High Speed PA scheme on FPGA}

%%从算法流程中可以发现，大数乘法是MMH和MH中的共同核心部分，所以本文设计了一种大数乘法可复用的方案结构及其配套的状态机，并详细介绍了关键参数的设定方法；随后，本文基于NTT设计了长度为768Kb的大数乘法模块；除此以外，一种低开销的模累加模块和流水的二进制模加模块也被设计来提高方案的整体性能。

A large-scale and high-speed PA scheme on the FPGA is designed based on MMH-MH PA algorithm in this section. It can be found that the major part of MMH-MH PA algorithm is the large-number multiplication according to Section 2. It is the core operation of both MMH function and MH function. Therefore, we designed a multiplication-reusable structure and its control unit for the scheme, and we introduced the calculation method of key parameters in this scheme. Afterwards, the design method of main units is illustrated. The most important unit is the large-number multiplication unit. It deeply determines the performance of the whole scheme. A 768Kb multiplication unit is designed based on number theoretic transform (NTT). In addition, it can optimize computation resource cost according to the real-time requirement by adjusting the radix of NTT. Then we designed a low-cost modular accumulation unit and a pipelined binary modular addition unit to improve overall performance. 
\subsection{Architecture of large-scale PA scheme}
\begin{figure}
	\includegraphics[height=4.8cm,width=9cm]{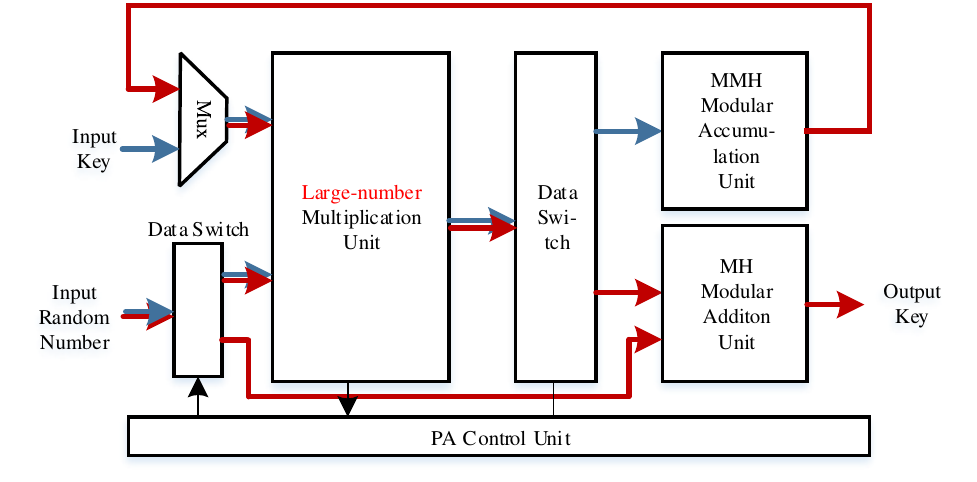}
	\caption{Architecture of large-scale PA scheme.  \textcolor{blue}{$\rightarrow$} means the data flow of MMH function and \textcolor{red}{$\rightarrow$} means the data flow of MH function.}
	\label{fig:2}       % Give a unique label
\end{figure}

The architecture of large-scale PA scheme is indicated as Fig. \ref{fig:2}. The multiplication unit is reused in this architecture to reduce the resource cost. There are two streams of data flow in this scheme, and they represent the MMH function data stream and MH function data stream. 

% Fig 2

\begin{figure}
	\includegraphics[height=4cm,width=6cm]{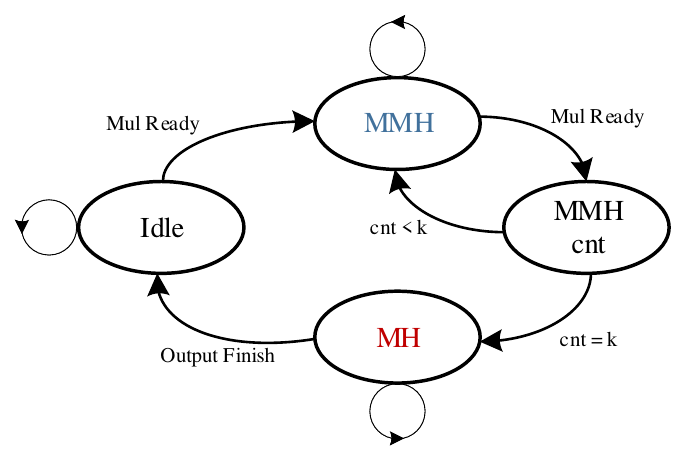}
	\caption{State diagram of control unit.}
	\label{fig:3}       % Give a unique label
\end{figure}

A matched control unit is designed to control computational process and data flow as indicated as Fig. \ref{fig:3}. The MMH function calculation begins first when the multiplication unit is ready. Because MMH function needs $k$ times multiplication, the state will turn to "MMH cnt" and a counter up one when one multiplication operation completes. The state will return to "MMH" if $cnt < k$ and go to "MH" if $cnt = k$. "MH" state will calculate MH function and output the final key, and the state will go back to initial state until the end of the output. The control unit will make data flow follow the blue arrow at the state "MMH" and the red arrow at the state "MH".

\subsection{Key parameter calculation of large-scale PA scheme}
%明确需要确定的参数：1. subblock size 2. subblock numbers
The sub-block size $\gamma$ and the sub-block number $k$ are the most critical parameter in this scheme. Because the input block size $n = \gamma \times k$ is the main optimization target, $\gamma$ and $k$ are expected to be as larger as possible. $\gamma$ is restricted by two conditions: 1. $2^\gamma-1$ should be a primer; 2. $2^\gamma-1$ should be less than the largest number $N_{mul}$ supported by the large-number multiplication unit. 
%确定subblock size的确定方法
The $N_{mul}$ of multiplication unit in our implementation scheme is $2^{786432}-1$, so the sub-block size $\gamma$ can be chosen as $756839$ ($2^{756839}-1$ is the 32nd Mersenne prime). The largest number multiplication unit on FPGA as we know is the module in \cite{Ye2018}, and its $N_{mul} = 2^{1179648}-1$. Therefore, the largest $\gamma$ can be $859433$ ($2^{859433}-1$ is the 33rd Mersenne prime). 
%确定subblock number的方法
The sub-block number $k$ is restricted by the compression ratio $R_{PA}$ of PA (The calculation method of $r$ can be referred in \cite{Bingze2021}), and $1/k$ should be larger than $R_{PA}$. The compression ratio $R_{PA}$ is affected by the specific QKD system and transmission distance, so the specific value of $k$ will be discussed in next section. 

Then we elaborate the design of three main modules of this scheme: 1. the large-number multiplication unit; 2. the low-cost modular accumulation unit 3. the pipelined binary modular addition unit. 

\subsection{Design of large-number multiplication unit}
%大数乘法模块的设计
The large number multiplication unit is the most essential and complex unit in this scheme. The size of this large number multiplication unit is $786432$ bits. It is implemented based on number theoretical transform (NTT) algorithm. The large number multiplication algorithm ($Z=X \times Y$) can be summarized as follow:

1) Break the large numbers $X$ and $Y$ into a sequence of words $x(n)$ and $y(n)$ using base $B$: $X = \sum{x_i \times B^i}$ and $Y = \sum{y_i \times B^i}$.

2) Compute the dot product of NTT results $NTT(X)$ and $NTT(Y)$: $Z'_i$ =  $NTT(X)_i \times NTT(Y)_i$.

3) Compute the inverse NTT (INTT) : $Z'' = INTT(Z')$.

4) Resolve the carries: let $Z''_{i+1} = Z''_{i+1} + Z''_i/B$, and $Z_i = Z'_i \textbf{mod} B$.

%重要参数

The base $B=24$ and the sequence size $n = 32768$, so the size of this large number multiplication is $n \times B = 786432$. NTT and INTT are the main parts in this algorithm, and a 65536-point NTT and 65536-point INTT are required. An N-point NTT is defined as:
\[{X_k} = \sum\limits_{n = 0}^{N - 1} {{x_n}{{({W_N})}^{nk}}\bmod p}. \]
And an N-point INTT is defined as:
\[{x_k} = {N^{ - 1}}\sum\limits_{n = 0}^{N - 1} {{X_n}{{({W_N})}^{ - nk}}\bmod p}. \]

To simplify the modulo operation, $p$ is chosen as a special primer, which is $p = 2^{64} - 2^{32} + 1$. One data point in NTT $NTT(X)_i$ is represented as a 64-bits digit. So the largest data point is $NTT(X)_i \times NTT(Y)_i$, which is a 128-bits digit and represented as $2^{96}a+2^{64}b+2^{32}c+d$. It can be rewrite as,
\[\begin{array}{c}
{2^{96}}a + {2^{64}}b + {2^{32}}c + d{\rm{           }}(\bmod p)\\
\equiv  - 1(a) + ({2^{32}} - 1)b + ({2^{32}})c + d\\
\equiv ({2^{32}})(b + c) - a - b + d\quad\quad\quad
\end{array}\]
 
%直接计算NTT的计算量较大，因此需要蝶形算法减小计算量。鲽形算法的基数是算法的重要参数，基数越大，算法的速率越快，但是消耗的计算资源越多，基数越小算法的速率越慢但是需要的计算资源越少。下面将简要介绍这种蝶形算法。

The computation complexity of directly computing NTT is too excessive, so the butterfly algorithm is required to reduce it. The radix is an important parameter of butterfly algorithm. Larger radix will decrease the run time of algorithm and cost more computational resource. The radix-$r$ butterfly algorithm will be introduced next.

%%不同基的蝶形算法
\subsubsection{Radix-$r$ butterfly algorithm}

We take 16-point NTT as an example to demonstrate the difference between radix-2, radix-4, radix-16 butterfly algorithm. The computation of radix-16 is indicated as follow,

\[{X_k} = \sum\limits_{n = 0}^{15} {{x_n}{{({W_{16}})}^{nk}}\bmod p} \].

Obviously, the radix-16 algorithm only needs to run once to complete the 16-point NTT. The computation of radix-4 is indicated as follow, 

\[{X_k} = \sum\limits_{n = 0}^{3} {{x_n}{{({W_{4}})}^{nk}}\bmod p} \].

The 16-point NTT can be divided into twice radix-4 calculation, the specific process is shown below,

\[\begin{array}{c}
{X_k} = \sum\limits_{n = 0}^{{\rm{16}}} {{x_n}{{({W_{{\rm{16}}}})}^{nk}}\,\bmod \,p} \\
{\rm{ = }}\sum\limits_{{n_{\rm{0}}} = 0}^{\rm{3}} {{{({W_4})}^{{n_0}{k_1}}}\left\{ {\sum\limits_{{n_1} = 0}^3 {{x_n}{{({W_4})}^{{n_1}{k_0}}}} } \right\}\,{{({W_{16}})}^{{n_1}{k_1}}}\bmod \,p} 
\end{array}\]

,where $n=4n_0+n_1$ and $k=4k_0+k_1$. In the same way, the 16-point NTT can be divided into quartic radix-2 calculation. 

\begin{figure}
	\includegraphics[height=2.5cm,width=6cm]{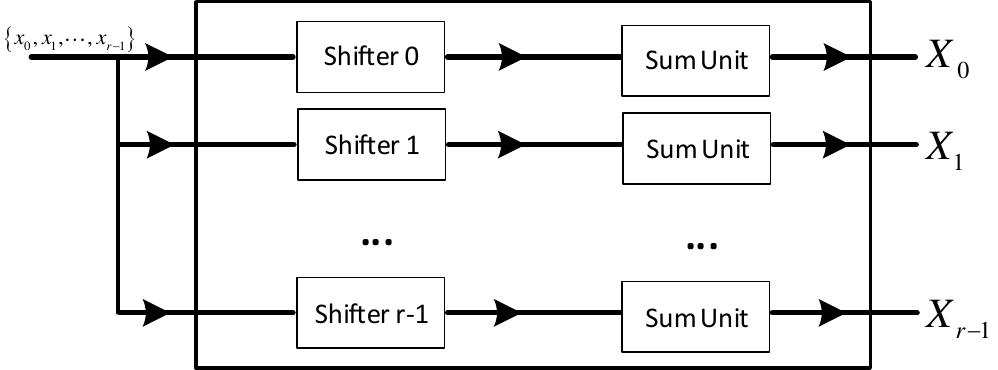}
	\caption{Structure of radix-$r$ unit.}
	\label{fig:4}       % Give a unique label
\end{figure}

It is worth noting that the rotation factor $W$ can be the power of 2 when $p = 2^{64} - 2^{32} + 1$, such as $W_{16} = 4096 = 2^{12}$. Then the multiplication can be replaced by shifting. A radix-r calculation structure can be indicated as Fig. \ref{fig:4}. It can be found that the unit with larger radix-$r$ costs more computation resource.

%为了评估该方案的实时性，我们选择了基16的单元来设计方案以获得最好的实时性

To evaluate the real-time of scheme, we choose the best real-time selection, radix-16, in this scheme. 

\subsubsection{structure of large-number multiplication unit}

\begin{figure}
	\includegraphics[height=8cm,width=10cm]{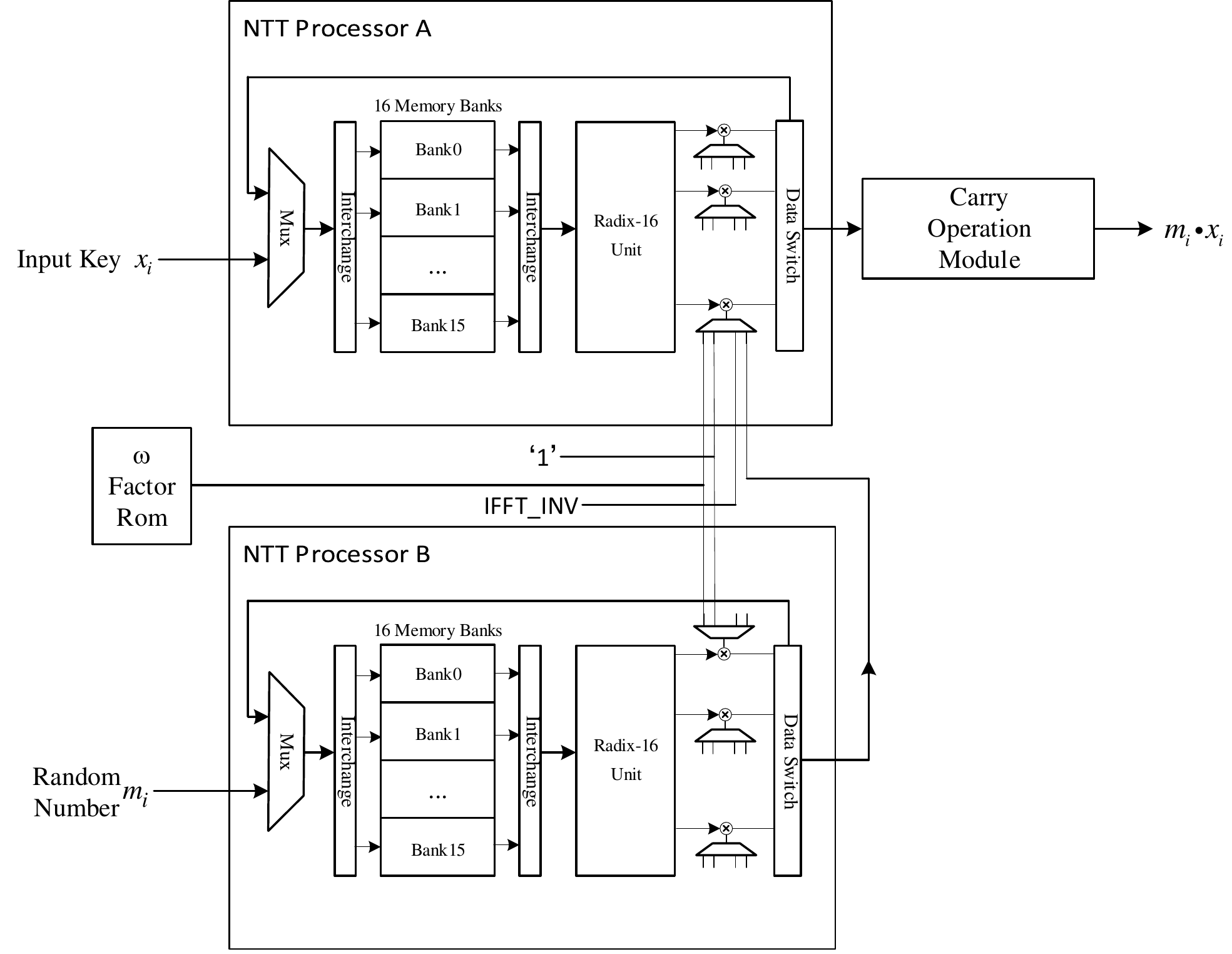}
	\caption{Structure of large-number multiplication unit with $radix=16$.}
	\label{fig:5}       % Give a unique label
\end{figure}

The structure of large-number multiplication unit is indicated as Fig. \ref{fig:5}. The NTT processor 
used a radix-16 unit and matched memory to complete a $16^4=65536$-point NTT/INTT calculation. The calculation requires four stage to complete. The memory unit is divided into 16 banks to load 16-point data in one time. The data in memory should be stored by a well-designed address mapping table. Details of the well-designed address mapping table can be referred in \cite{Wang2014}.

The data is load into memory before the NTT calculation. In each stage of NTT, the data is access and transmit into the radix-16 unit. Then it is send to a 64-bit multiplication unit, and the multiplicand will be constant '1', rotation factor $W_N^k$, INTT factor $N^{-1}$  and the NTT results of NTT\_B. The $65536$th primitive root $W_{65536}$ of $p=2^{64}-2^{32}+1$ is 0x$ed3365469864f124$. %0x$1467953c5cd19ec5$
After NTT and INTT calculation are completed, the data is load into the carry option module to guarantee each point of multiplication results is 24bits. More details of large-number multiplication unit can be found in \cite{Wang2014}.

\subsection{Design of low-cost modular accumulation unit}

The low-cost modular accumulation unit is responsible for modular accumulate calculation in MMH function $y = \sum\limits_{i=1}^k {{y_i}\bmod p} $, where $y_i$ is the multiplication result of large-number multiplication unit and $p = 2^{756839} - 1$. The modular addition can be simplified as follow, 

\[\begin{array}{c}
a + b\bmod ({2^{756839}} - 1)\\
= a\bmod {2^{756839}} + \left\lfloor {a/{2^{756839}}} \right\rfloor \\
+ b\bmod {2^{756839}} + \left\lfloor {b/{2^{756839}}} \right\rfloor 
\end{array}\]

\begin{figure}
	\includegraphics[height=3.2cm,width=8cm]{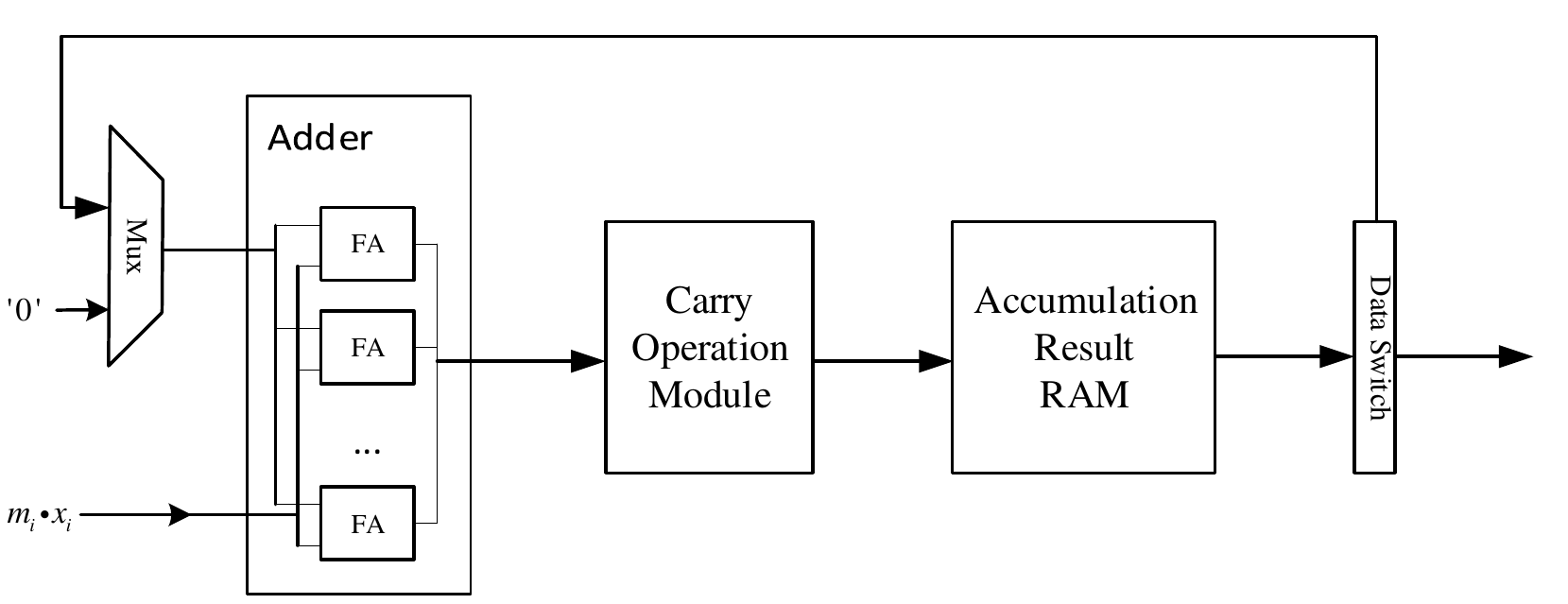}
	\caption{Structure of low-cost modular accumulation unit.}
	\label{fig:6}       % Give a unique label
\end{figure}

In this way, the modular calculation is replaced by addition and bit operation. It only needs full adders and $756839$ bits memory. The structure is indicated as Fig. \ref{fig:6}. In this structure, the input data just  adds the data in accumulation result memory with period 756839 and clears memory when once MMH function completes. 

\subsection{Design of pipelined binary modular addition unit}

\begin{figure}
	\includegraphics[height=3.2cm,width=8cm]{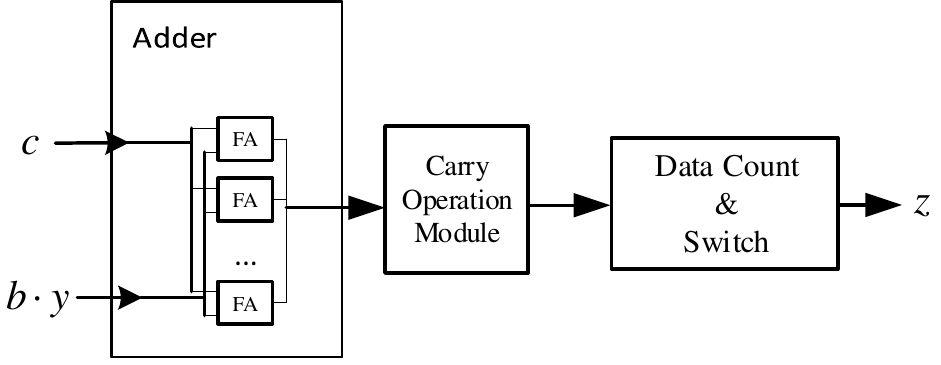}
	\caption{Structure of pipelined binary modular addition unit.}
	\label{fig:7}       % Give a unique label
\end{figure}

%需要计算的公式和总体结构
The pipelined binary modular addition unit is designed to calculate the equation 
$z = (b \cdot y + c\bmod {2^\alpha })/{2^{\alpha  - \beta }}$. The structure of pipelined binary modular addition unit is indicated as Fig.~\ref{fig:7}. $b \cdot y$ has been prepared by multiplication unit and is the unit input. adder and carry operation module calculate $b \cdot y + c$. The binary modular and division is implemented by the data counter and switch module. The parameter $\alpha$ is equal to $\gamma$ and $\beta$ is equal to the length of secure key. Each frame of data is 24 bits, so the module begins to output data when the input data count is $\left\lfloor {{{(\alpha - \beta)} \mathord{\left/{\vphantom {{( \alpha - \beta )} {24}}} \right.\kern-\nulldelimiterspace} {24}}} \right\rfloor $. The first frame of data outputs $\left( {  \alpha - \beta} \right)\bmod 24$ bits data. The rest of frames are 24bits each frame. The output ends when the data count is $\left\lfloor {\alpha /24} \right\rfloor$. This unit use data counter instead of calculation module implementing pipelining and low cost.

\section{Implementation and Experiment}
%总起：
The proposed large-scale and high-speed (LSHS) PA scheme is implemented on the Xilinx ZYNQ ultrascale+ evaluate kit. This kit is based on a Zynq Ultrascale+ XCZU9EG-2FFVB1156 FPGA. The resource utilization is an important indicator to estimate the practicability of a FPGA-based PA scheme, and it is influenced by $\gamma$ and $radix$ in our scheme. The resource utilization  of our scheme with $\gamma = 756839$ and $radix = 16$ is indicated as Table. \ref{tab:1}.

\begin{table}
	\caption{The resource utilization of LSHS PA scheme\tnote{1}}
	\label{tab:1}       % Give a unique label
	\begin{threeparttable}          %这行要添加
		\begin{tabular}{cccc}
			\hline
			\hline
			Resource & Scheme Used & Available\tnote{1} & Utilization Rate \\
			\hline
			\hline
			Luts & 156707 & 230400 & 68\% \\
			BRAMs & 198 & 408 & 48\% \\
			DSP Slices & 512 & 1728 & 30\% \\
			\hline
			\hline
		\end{tabular}
		\begin{tablenotes}    %这行要添加， 从这开始
			\footnotesize               %这行要添加
			\item[1] Available: Zynq Ultrascale+ XCZU9EG-2FFVB1156 available       %这行要添加
			\item[2] $\gamma = 756839$ and $radix = 16$
		\end{tablenotes}            %这行要添加
	\end{threeparttable}          %这行要添加
\end{table}

%该方案进行了三方面的测试

This scheme is evaluated in three aspects: 1. The input block size of the LSHS PA scheme is evaluated and the secure key rate improvement of a FPGA-based QKD system with LSHS PA scheme is demonstrated; 2. The throughput of the LSHS PA scheme on different block size is evaluated and compared with existing PA schemes; 3. The resource cost and core index of the LSHS PA scheme are compared with existing FPGA-based PA schemes.
 
%第一部分需要展示方案能达到的最大码长和安全码率：
\subsection{The input block-size and secure key rate of the LSHS PA scheme}
%输入码长的计算
The input block size $n$ of the LSHS PA scheme is equal to $k \times \gamma$. $\gamma$ is a fixed value and $k$ is related to the maximum compression ratio $R^{MAX}_{PA}$ of PA. Therefore, the input block size $n$ is related to the maximum compression ratio $R^{MAX}_{PA}$ of PA. $R_{PA}$ can be calculated on the basis of QKD system parameters. The compression ratio of a DV-QKD system can be calculated by $R_{PA} = \beta {I_{AB}} - {I_{AE}(e_1+\Delta_n)} $. The compression ratio of a CV-QKD system can be calculated by $R_{PA} = \beta {I_{AB}} - {\chi _{BE}} - \Delta_n$. The main fluctuate factor of compression ratio is the channel error rate $e$. The maximum compression ratio $R^{MAX}_{PA}$ means the compression ratio $R_{PA}$ when the channel error rate $e$ is minimum. The parameter $k$ is required to be smaller than $1/{R^{MAX}_{PA}}$ to maximum the secure key rate.
%实验结果展示

\begin{figure}
	\includegraphics[height=13cm,width=9cm]{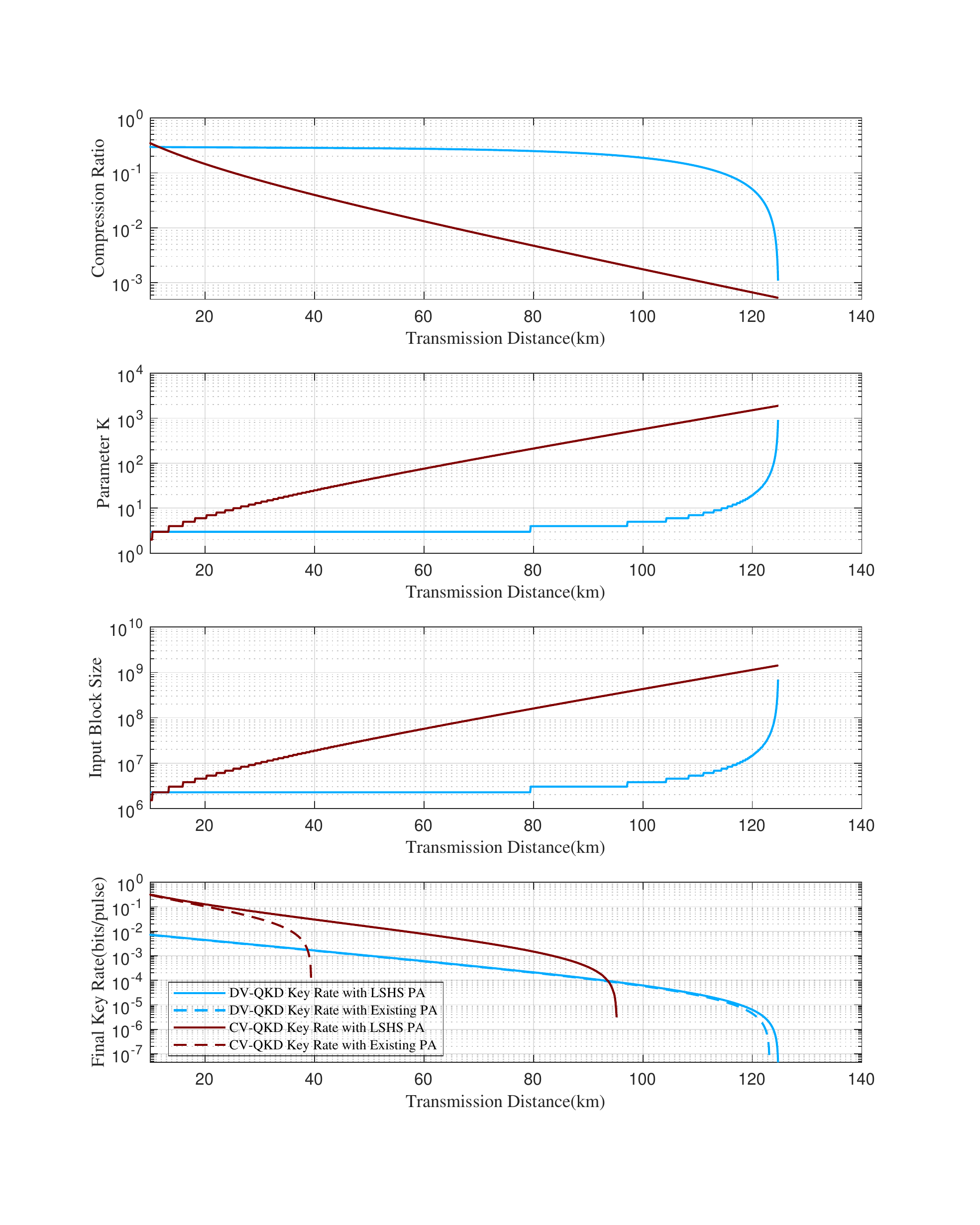}
	\caption{Final key rate of a QKD system with LSHS PA.}
	\label{fig:8}       % Give a unique label
\end{figure}

The effect of the LSHS PA scheme on the input block size and secure key rate is evaluated by the simulation on a typical DV-QKD system~\cite{Yuan2018} and a typical CV-QKD system \cite{Zhang2019b}. The results demonstrate the parameter $k$, the input block size and secure key rate as communication distance changes in Fig. \ref{fig:8}. We computed the compression ratio of PA $R_{PA}$ in two QKD systems at different transmission distances to confirm the parameter $k$ in the first subgraph. Here we assumed the input block size is infinite to compute the maximum compression ratio in the infinite input block size case. The parameter $k$ of the LSHS PA scheme was calculated according to the compression ratio of PA $R_{PA}$ in the second subgraph. Then the input block size of the LSHS PA scheme at different transmission distances was confirmed with the parameter $k$ and the parameter $\gamma = 756839$ in the third subgraph, where the input block size $N$ equals to $k \times \gamma$. Finally, we simulated the final key rate of typical QKD systems with the LSHS PA scheme at different transmission distances in the last subgraph. The input block size of existing FPGA-based PA schemes is set to $10^6$, because it is the largest input block size of existing FPGA-based PA schemes as we know. It can be found that the effect of the LSHS PA scheme is more significant in the CV-QKD system. This is because the finite size effect in the CV-QKD system is more serious than that in the DV-QKD system. 

%实验结论
In conclusion, the simulation results indicate that the LSHS PA scheme can improve the input block size and the secure key rate of a QKD system. This improvement is more efficient in a CV-QKD system. 

% 第二部分需要展示不同码长下的处理速率：
\subsection{The throughput of the LSHS PA scheme}

\begin{figure}
	\includegraphics[height=5cm,width=9cm]{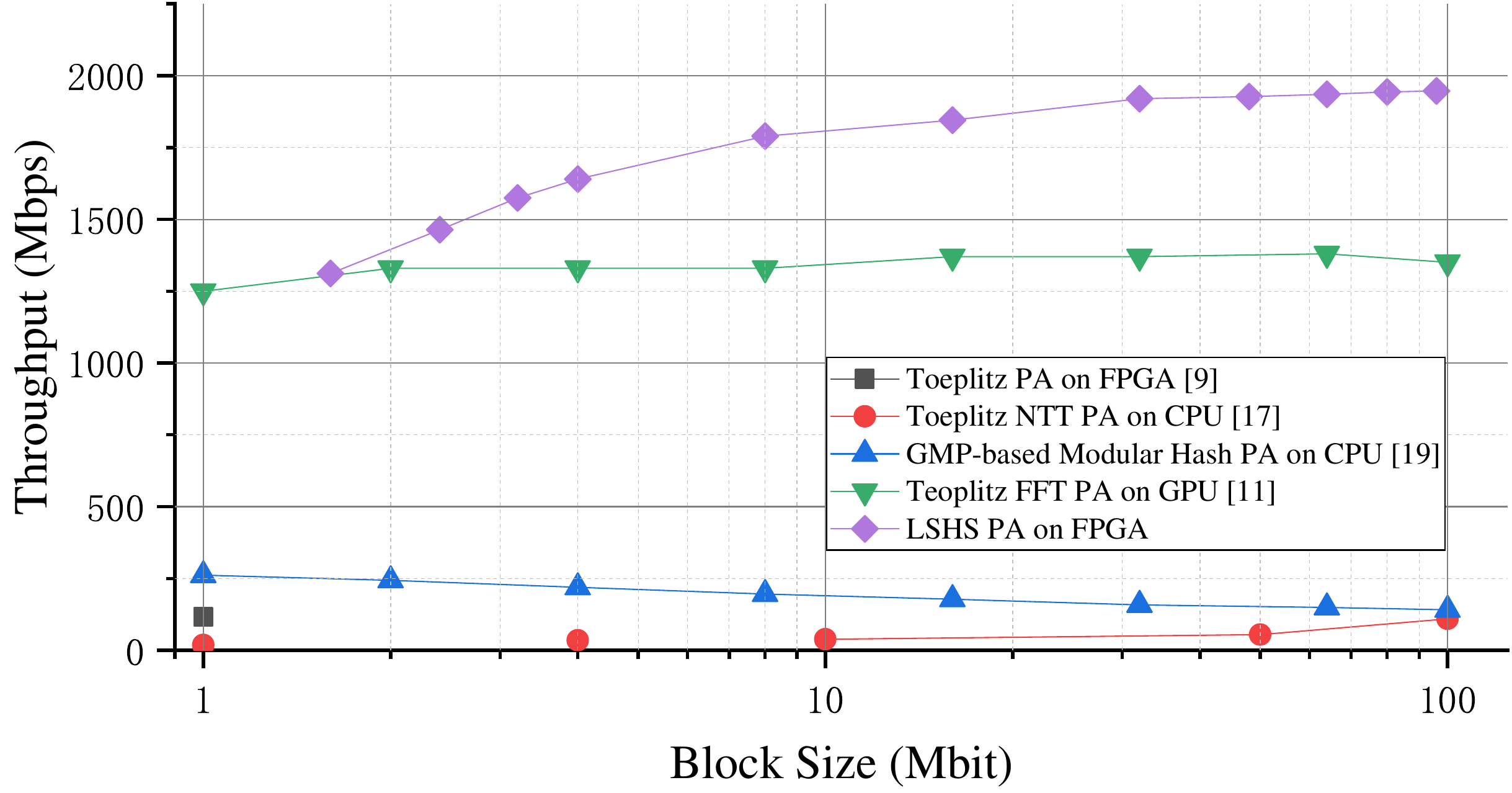}
	\caption{The throughput comparison of existing FPGA-based PA schemes}
	\label{fig:9}       % Give a unique label
\end{figure}

The throughput of PA means the maximum rate of the input key into a PA scheme. It is an important index of PA, which affects real-time of a QKD system. We experimented the throughput of the LSHS PA scheme at different input block sizes with a random simulated data source. Then, we compared the throughput between our scheme and existing schemes as indicated in Fig. \ref{fig:9} \cite{Li2019a,XiangyuWangYichenZhangSongYu2016a,Yuan2018,Yan2020}. 

The comparison shows that our scheme improves the throughput by an order of magnitude than existing FPGA-based PA schemes, and our scheme reaches a equal throughput with the existing best throughput PA scheme -- a GPU-based PA scheme. The throughput advantage of the LSHS PA scheme profits from two reasons: 1. the less computation of MMH-MH PA algorithm; 2. the three efficient units we design in section 3.   

%第三部分资源消耗对比
\subsection{The resource cost and core index comparison of FPGA-based PA schemes}

We compared the resource cost and core index between LSHS PA scheme and existing FPGA-based PA schemes. The resource cost of a FPGA-based PA scheme includes look up tables (LUTs), internal Random Access Memory (RAM), external RAM and DSP slices. The core index to be compared is throughput and input block size of a PA scheme. The comparison results are indicated as Table \ref{tab:2}. 

The LSHS PA scheme costs the most LUT and Internal-RAM among existing PA schemes. The main cost of LSHS PA is the large-number multiplication unit. It costs nearly $90\%$ of LSHS PA scheme. The main reason of its high cost is that it chooses the radix of the unit as $16$. This provides the best real-time and throughput, but also costs the most resource. If a large-number multiplication unit with $radix=2$ is used in this scheme, we estimated that the throughput and the LUTs cost will be reduced to approximately eighth of their previous. The internal-RAM cost of the LSHS scheme is the most, because it uses the NTT algorithm for acceleration, and the NTT algorithm needs to store all the input data and rotation factors. Similarly, the scheme in \cite{Li2019a} used the FFT algorithm for acceleration. The NTT algorithm has an advantage over FFT algorithm, that the data format of the NTT is the integer instead of the floating-point of the FFT. This advantage significantly reduces the memory cost of the LSHS scheme compared with the scheme in \cite{Li2019a}, and frees the LSHS PA scheme from dependence on the external-RAM. 

\begin{table}
	\caption{The resource cost and core index comparison of FPGA-based PA schemes}
	\label{tab:2}       % Give a unique label
	\begin{threeparttable}          %这行要添加
		\begin{tabular}{ccccc}
			\hline
			\hline
			& Yang et al. & Constantin et al. & Li et al. &  \makecell[c]{LSHS PA}  \\
			\hline
			\hline
			LUTs & 15,604 & 26,571 & 37,203 & 156,707 \\
			Internal-RAM & 100Kb & 0Kb & 5,652Kb & 11,232Kb \\
			External-RAM & 0Kb & 1,095Kb & 128Mb & 0Kb \\
			Throughput & 64Mbps & 41Mbps & 116Mbps & 1,400Mbps \\
			Input Block Size & 1Mb & 1Mb & 1Mb & 1-1000Mb \\ 
			\hline
			\hline
		\end{tabular}
	\end{threeparttable}          %这行要添加
\end{table}

Although the LSHS PA scheme costs more resources, the LSHS PA scheme greatly improves the core index of FPGA-based PA scheme. Before the LSHS PA scheme, the throughput and input block size of the FPGA-based PA scheme is far below that of PA schemes based on other platforms. Therefore, although FPGA-based PA schemes have advantages of low power consumption and high integration level for QKD systems, they are not applied widely due to the low core index. The core index of the LSHS PA scheme has exceeded existing FPGA-based PA schemes by several orders of magnitude, and it has surpassed the best PA scheme in term of these core indexes.       
\section{Conclusion}

In this research, a large scale and high speed PA scheme based on FPGA is proposed to improve the core index (input block size and throughput) of a FPGA-based PA scheme. This scheme is designed based on the MMH-MH PA algorithm. We designed the architecture of the LSHS PA scheme to reuse as many computation unit as possible for the resource cost reduction. Then we focused on the design of three key units of this scheme in this paper. A 786432-bits large-number multiplication unit is designed based on the NTT algorithm. A low-cost modular accumulation unit is designed to compute $mod~2^{756839}-1$ accumulation with minimal memory. A pipelined binary modular addition is designed to compute the binary modular of arbitrarily length without memory. We implemented the proposed large-scale and high-speed PA scheme on the Xilinx ZYNQ ultrascale+ evaluate kit. We referenced the parameters of a typical DV-QKD system and a typical CV-QKD system to evaluate the input block size, the throughput and the influence on the final key rate of the LSHS PA scheme. We compared these results with existing PA schemes on FPGA and other platforms. The results indicate that the LSHS PA scheme has improved the throughput by an order of magnitude, and it can improve the input block size by several magnitudes compared with existing FPGA-based PA schemes. The input block size improvement is more obvious in a CV-QKD system and a long transmission distance QKD system. The above results indicates that the LSHS PA scheme can significantly improve the final key rate of a FPGA-based QKD system. It is worth noting that the core index of the LSHS PA scheme has exceed the existing best PA scheme of all platforms. Adding the consideration with the power consumption and integration level advantages of the FPGA-based PA schemes, the LSHS PA scheme is a highly competitive solution for the QKD systems.

%加一段平台的对比，说明该方案是全平台最有竞争力的方案

% or


%%\appendices
%%\section{Proof of the First Zonklar Equation}
%%Appendix one text goes here.

%%\section{}
%%Appendix two text goes here.

%%\section*{Acknowledgment}

%%\ifCLASSOPTIONcaptionsoff
%%  \newpage
%%\fi



\bibliographystyle{IEEEtran}

\bibliography{library}
%\begin{thebibliography}{1}


%\end{thebibliography}


%%\begin{IEEEbiography}{Michael Shell}
%%Biography text here.
%%\end{IEEEbiography}

%%\begin{IEEEbiographynophoto}{John Doe}
%%Biography text here.
%%\end{IEEEbiographynophoto}

%\newpage

%%\begin{IEEEbiographynophoto}{Jane Doe}
%%Biography text here.
%%\end{IEEEbiographynophoto}


%\vfill


\end{document}